\documentstyle[aas2pp4,graphics]{article} 
\slugcomment{Submitted to Astrophysical Journal}
\lefthead{Morsink, Stergioulas and Blattnig} 
\righthead{Quasi-normal modes of rotating relativistic stars...} 
 
\begin{document} 
 
\title{Quasi-normal modes of rotating relativistic stars - neutral modes 
for realistic equations of state}


\author{Sharon M. Morsink\altaffilmark{1}, 
Nikolaos Stergioulas\altaffilmark{2}, and Steve R. Blattnig\altaffilmark{1}}

\altaffiltext{1}{Department of Physics, University of Wisconsin-Milwaukee, \\ 
PO Box 413, Milwaukee, WI 53201, USA \\ 
{\rm morsink@pauli.phys.uwm.edu}; {\rm srb2@csd.uwm.edu}}
\altaffiltext{2}{Max Planck Institute for Gravitational Physics (Albert-Einstein-Insitute) \\ 
 D-14473 Potsdam, Germany\\ 
{\rm niksterg@aei-potsdam.mpg.de}}
 
\begin{abstract} 
   
We compute zero-frequency (neutral) quasi-normal $f$-modes of fully 
relativistic and rapidly rotating neutron stars, using several realistic 
equations of state (EOSs) for neutron star matter. The zero-frequency modes 
signal the onset of the gravitational radiation-driven instability. We find 
that the $l=m=2$ (bar) $f$-mode is unstable for stars with gravitational mass 
as low as $1.0 - 1.2M_\odot$, depending on the EOS.  For $1.4M_\odot$ neutron 
stars, the bar mode becomes unstable at $83 \% -93 \%$ of the maximum allowed 
rotation rate.  For a wide range of EOSs, the bar mode becomes unstable at a 
ratio of rotational to gravitational energies  $T/W \sim 0.07-0.09$ for 
$1.4 M_\odot$ stars and $T/W \sim 0.06$ for maximum mass stars. This is to be 
contrasted with the Newtonian value of $T/W \sim 0.14$.  We construct the 
following empirical formula for the critical value of $T/W$ for the bar mode, 
$(T/W)_2 = 0.115 -0.048 \;M / M_{\rm max}^{\rm sph}$, which is insensitive to 
the EOS to within $4-6\%$. This formula yields an estimate for the neutral mode
sequence of the bar mode as a function only of the star's mass, $M$, given the
maximum allowed mass, $M_{\rm max}^{\rm sph}$, of a nonrotating neutron star. 
The recent discovery of the fast millisecond pulsar in the supernova remnant 
N157B,  supports the suggestion that a fraction of proto-neutron stars are born
in a supernova collapse  with very large initial angular momentum. If some 
neutron stars are born in an accretion-induced-collapse of a white dwarf, then
they will also have very large angular momentum at birth. Thus, in a fraction 
of newly born neutron stars the instability is a promising source of continuous
gravitational waves. It could also play a major role in the rotational 
evolution (through the emission of angular momentum) of merged binary neutron 
stars, if their post-merger angular momentum exceeds the maximum allowed to 
form a Kerr black hole.

\end{abstract} 


\section{Introduction} 
\label{s:introduction} 
 
A core-collapse supernova or the accretion-induced-collapse of a white 
dwarf can result in the birth of a hot, rapidly rotating neutron star. 
During the first year of its life (while it cools from $\sim 10^{10} - 
10^{9}$K) the neutron star will be unstable to the emission of 
gravitational radiation due to the Chandrasekhar-Friedman-Schutz (CFS) 
nonaxisymmetric instability (Chandrasekhar 1970; Friedman \& Schutz 
1978; Friedman 1978). The instability will only be operating while the 
star is rotating more rapidly than some critical angular velocity. 
Via the instability, gravitational waves carry away a significant 
amount of the star's angular momentum.  This early spin-down epoch has 
two important astrophysical implications: First, the gravitational 
radiation emitted may be detectable by the planned gravitational wave 
detectors. (Note that this discussion is also relevant to post-merger 
objects in a neutron star binary coalescence.)  Second, it may be 
possible to indirectly observe the critical angular velocity through 
the detection of young, rapidly rotating pulsars in supernova 
remnants, such as PSR J0537-6910 (Marshall et al. 1998). 
 
The critical velocity for the onset of the CFS-instability  in 
polar perturbations, ($f$-modes) has been 
computed before in various approximations: in the Newtonian limit 
(Managan 1985; Imamura, Friedman, \& Durisen 1985; Ipser \& Lindblom 
1990), the post-Newtonian approximation (Cutler 1991; Cutler \& 
Lindblom 1992; Lindblom 1995), and the relativistic Cowling 
approximation for polytropes (Yoshida \& Eriguchi 1997) and 
realistic equations of state (EOSs) (Yoshida \& Eriguchi 1998). 
For a detailed review, see 
Stergioulas (1998).  The first fully-relativistic computation of the 
onset of the instability in $f$-modes is presented in Stergioulas 
(1996) and Stergioulas \& Friedman (1998, hereafter SF). SF find a 
gauge in which six perturbed field equations can be solved 
simultaneously on a finite grid with good accuracy. Using polytropic 
equations of state with index $N=1.0$, 1.5 and 2.0, SF show that 
general relativity has a significant effect on the onset of the 
instability, lowering the rotation rate at which it occurs, as the 
star becomes more relativistic. A surprising result is that the 
$l=m=2$, ``bar'' $f$-mode  instability (which in the Newtonian 
limit exists only for stiff polytropes of index $N<0.808$) exists for 
relativistic polytropes with index as large as $N=1.3$. SF suggested 
that the $l=m=2$  instability should also exist for realistic 
EOSs, which is confirmed in the present paper. In the 
Newtonian limit the gravitational-wave-driven and viscosity-driven 
 bar mode instabilities occur at the same value 
of the ratio of rotational energy to 
the gravitational binding energy, $T/W\sim0.14$. 
SF conjectured that when effects due to relativity 
are included, the onset of the two types of instabilities will be 
split, with the CFS instability occurring at lower values of $T/W$ 
and the viscosity-driven instability at higher values. 
 Calculations of relativistic effects on the viscous instability 
  (Shapiro \& Zane 1997; Bonazzola, Frieben, \& Gourgoulhon 1998) 
  agree with this conjecture. 
 
For a perturbation with azimuthal angular dependence $e^{im\phi}$, 
modes with the smallest value of  the spherical harmonic multipole 
index $l$ will have the fastest growth rate 
and the highest gravitational radiation luminosity. Hence the modes with 
$l=m$ and in particular, with $m=2$, are the most relevant for 
astrophysics.  For a perfect fluid all modes with $m\ge2$ are of 
interest, however, when imperfect fluid effects are included, (Cutler 
and Lindblom 1987; Ipser and Lindblom 1991; Lindblom 1995; Yoshida \& 
Eriguchi 1995) polar modes with $ m>5$  will always be damped by 
  shear and bulk viscosity. 
 
In the present paper, we use the SF  scheme 
to determine the onset of the CFS instability of $f$-modes with $l=m$ 
and $2\leq m\leq 5$ for realistic EOSs. We also improve 
on the numerical implementation of the method, by using a new 
finite-difference scheme in the angular direction and an improved 
algorithm for locating the exact onset of the instability with higher 
accuracy.  We find that the realistic EOSs show similar behaviour as 
the polytropic EOSs in SF. The $l=m=2$ $f$-mode becomes unstable for 
all realistic EOSs examined, for stars with masses as low as $M=1.0 - 
1.2 M_\odot$, depending on the EOS.  Stars with mass near $1.4 M_\odot$
are unstable to the bar mode at $83\% - 93\%$ of the mass-shedding 
(Kepler) limit.

As was first noticed by Andersson (1998), the critical angular 
velocity for axial $r$-modes, in a perfect fluid star, is exactly 
zero, so that all stars are generically unstable for all values of $m$ 
(Friedman \& Morsink 1998).  Again, the inclusion of viscosity will 
stabilize all modes except those with the lowest values of $m$. Two 
independent computations including the effects of viscosity in 
Newtonian stars (Lindblom, Owen \& Morsink 1998; Andersson, Kokkotas 
\& Schutz 1998) estimate that the lowest angular velocity for which 
$l=m=2$ \ $r-$mode is unstable is roughly $6\% - 20\%$ of the 
Kepler limit for uniformly rotating stars. 
 
The following scenario may describe the early spin-evolution of a 
newly born neutron star, if it is born with an angular velocity close 
to the Kepler limit $\Omega_K$. While the star cools from $\sim 
10^{10}$K to $10^9$K, viscous effects will be small enough that the 
gravitational radiation instability will spin down the star.  In 
this temperature window the spin evolution will go through two phases. 
In the first phase, the star is rotating fast enough that both $f-$ 
and $r-$modes will be unstable.  During the second phase, only the 
$r-$modes are unstable. The determination of which mode will be the 
dominant mechanism for the shedding of angular momentum during the 
first phase will depend on the relative growth times for both types of 
modes. At present, the growth times for neither type of mode have been 
determined for rapidly rotating relativistic stars. 
 
The plan of this paper is as follows. In section 
\ref{s:nonaxisymmetric} we briefly review the method for computing the 
onset of the polar-mode nonaxisymmetric instability in relativistic 
stars. In section \ref{s:improved} we present the improvements in the 
numerical implementation of the scheme. The equations of state 
selected are discussed in section \ref{s:eos}. In section 
\ref{s:results} we present the critical angular velocities for f-modes 
with $2\le m \le 5$ for a variety of equations of state. 
Astrophysical implications will be discussed in the concluding 
section.

\section{Nonaxisymmetric Perturbations} 
\label{s:nonaxisymmetric} 
 
\subsection{Quasi-normal Modes and the Onset of Instability} 
 
Taking advantage of the axisymmetry and stationarity of the 
equilibrium star, a general linear perturbation can be written as a 
sum of quasi-normal modes, characterized by the spherical harmonic 
indices $(l,m)$. In this way perturbations of scalars, such as the  
energy density 
can be analyzed as 
\begin{equation} 
 \delta \epsilon = \epsilon_l(r) P_l^m(\cos \theta) e^{i(\omega_i t +m \phi)}, 
\end{equation} 
 where $P_l^m(\cos \theta)$ are the Legendre functions and 
  $\omega_i$ is the frequency of the mode in the inertial frame. 
 Perturbations of vector quantities, such as the four-velocity, 
  can be written in terms of vector harmonics, while the perturbation 
  in the metric can be written in terms of scalar, vector and tensor 
  harmonics (see Regge and Wheeler 1957). Vector and tensor harmonics 
are of two types - polar, which transform as $(-1)^l$ under a parity 
transformation (under the combination of reflection in the equatorial 
plane and rotation by $\pi$) and axial, which transform as 
$(-1)^{l+1}$ under parity. The  angulars part of polar vector 
harmonics are proportional to gradients of the spherical 
harmonics, while axial vector harmonics are proportional to the 
curl of a radial vector and a polar vector harmonic.

In the spherical limit, nonaxisymmetric perturbations decouple into 
purely polar and purely axial modes with unique values of 
$m$ and $l$. In a fluid, polar modes correspond to the $f$-, $p$- 
  and $g$- modes modes in the Newtonian limit, while axial modes 
correspond to $r$-modes in a Newtonian star (Papaloizou and Pringle 
1978). 
 
In a rotating star, the spherical symmetry is broken. While a mode  
can still be specified by a single value of $m$, the mode will no longer  
consist of a single $l$ harmonic. A polar $(l,m)$ 
mode acquires higher order, in $l$, polar terms, due to the 
non-sphericity of equipotential surfaces, and $(l \pm 1,m)$ and higher 
order, in $l$, axial terms, due to the coupling between polar and 
axial terms: 
\begin{equation} 
         P_l^{rot} \sim \sum_{l'=0}^\infty(P_{l+2l'} +A_{l+2l' \pm 1}). 
\end{equation} 
Similarly, an axial mode in a rotating star is written as a sum of 
axial and polar terms: 
\begin{equation} 
         A_l^{rot} \sim \sum_{l'=0}^\infty(A_{l+2l'} +P_{l+2l' \pm 1}). 
\end{equation} 
Thus, a normal mode of oscillation in a rotating star is defined as 
polar, if it reduces to a purely polar mode in the nonrotating limit 
and similarly for axial modes.  
 
Gravitational radiation drives a polar or axial mode of oscillation 
unstable, whenever the star rotates fast enough, that a perturbation 
which counter-rotates in the star's rest frame, appears to co-rotate 
with respect to a distant, inertial, observer.  Conservation of 
angular momentum dictates that the mode's angular momentum must 
decrease.  However, a counter-rotating perturbation's angular momentum 
(invariantly defined in the rotating frame), is negative, so that 
gravitational radiation causes a negative angular momentum 
perturbation to become more negative. For a given value of $m$, the 
instability first sets in via an $l=m$ mode, when the frequency of the 
mode vanishes in the inertial frame. Thus, the problem of finding the 
critical angular velocity reduces to finding solutions of the 
time-independent (zero-frequency) perturbation equations. 
 
\subsection{Solving for Time-independent Perturbations}

In this paper we will follow the method of Stergioulas \& Friedman 
(1998), where one can find a detailed presentation. Here we will 
only briefly sketch the solution method. In section \ref{s:improved} 
we will summarize improvements in the numerical implementation. 
 
In writing the perturbation equations for an axisymmetric and 
stationary, relativistic star, the Eulerian approach is followed (see 
Ipser \& Lindblom 1992 and Friedman \& Ipser 1992).  The Eulerian 
perturbations in the metric tensor, $\delta g_{ab} \equiv h_{ab}$, 
energy density $\delta \epsilon$ and four-velocity $\delta u^a$ are 
obtained by solving the system of equations consisting of the 
perturbed field equations and the perturbed equation of conservation 
of the stress-energy tensor: 
\begin{equation} 
   \delta R_{ab} = 8 \pi \bigl( \delta T_{ab} - \frac{1}{2} g_{ab} \delta T - 
                      \frac{1}{2} h_{ab} T \bigr), \label{dR} 
\end{equation} 
\begin{equation} 
  \delta ( \nabla_a T^{ab}=0), \label{dnT} 
\end{equation} 
where a perfect fluid stress-energy tensor 
\begin{equation} 
  T_{ab} = (\epsilon+P)u^au^b +P g_{ab}, \label{Tab} 
\end{equation} 
is assumed. In equation (\ref{Tab}), $\epsilon$ is energy density, $P$ is 
pressure and $u^a$ is the four-velocity of the fluid. 
 
Since linear perturbations are subject to a gauge freedom, only six 
components of the perturbed field equations need to be solved. The 
projection of equation (\ref{dnT}) normal to the four-velocity of the fluid 
i.e. the perturbed Euler equations, can be solved analytically for 
$\delta u^a$.  Next, one defines a function 
\begin{equation} 
  \delta U  = u_a \delta u^a  + \frac{1}{2} u^bu^ch_{bc}, 
\end{equation} 
so that the perturbation in the energy density becomes    
\begin{equation} 
  \delta \epsilon = \frac{(\epsilon +P)^2}{P \Gamma}  \bigl  
     ( \delta U + {1 \over 2} u^a u^b  
             h_{ab} \bigr ), \label{de}  
\end{equation} 
where 
\begin{equation} 
   \Gamma = \frac{\epsilon +P}{P} \frac{dP}{d \epsilon}  
\end{equation} 
is the adiabatic index of the perturbation (assumed to be equal to the 
adiabatic index of the equilibrium fluid). 
 
Thus, a zero-frequency mode is obtained by solving six components of 
the perturbed field equations (\ref{dR}) for $h_{ab}$ and the perturbed energy 
conservation equation 
\begin{equation} 
  \delta ( u_b \nabla_a T^{ab}=0), \label{dEn} 
\end{equation} 
for $\delta U$.  
SF found that by choosing the gauge as 
\begin{eqnarray} 
   h_{r \theta} &=& 0, \label{g1}\\ 
   h_{\theta \phi} &=& 0,\\ 
   h_{t \phi} &=& - \omega h_{\phi \phi},\\ 
   h_{\phi \phi} &=& \frac{h_{\theta \theta}}{r^2}e^{2(\psi-\alpha)}, \label{g4} 
\end{eqnarray} 
six components of the perturbed field equations can be solved 
simultaneously on a finite grid for the required boundary 
conditions, given a trial function for $\delta U$ that is close to  
its actual solution. 
 
The remaining equation (\ref{dEn}) is solved by expanding the function  
$\delta U$ in terms of suitably chosen basis functions $\delta U_i$ 
\begin{equation} 
   \delta U = \sum_i  a_i \delta U_i. \label{dU_exp} 
\end{equation} 
For polar modes, the basis functions are chosen to be 
\begin{equation} 
 \delta U_i = \delta U^{(jk)}_i = r^{l+2(j+k)} Y_{l+2k}^m (\cos \theta),  
\end{equation} 
obtained by letting $j$ and $k$ take different values $\geq0$ for each  
value of $i$. 
Equation (\ref{dEn}) is an equation  linear in $\delta U$ 
and can be represented schematically as 
\begin{equation} 
  L(\delta U) = 0, \label{L} 
\end{equation} 
where $L$ is the linear operator defined in Appendix C of SF.  
Substituting the expansion (\ref{dU_exp}) in equation (\ref{L}) 
and defining the inner product 
\begin{equation} 
 < \delta U_j | L | \delta U_i > = \int i \frac{\delta U_j}{m \Omega u^t}  
 L(\delta U_i) \sqrt{-g} d^3x, 
     \label{innerproduct}  
\end{equation} 
where $\Omega$ is the angular velocity of the star, a solution for  
the homogeneous equation (\ref{L}) exists, only when the determinant  
of the inner product matrix vanishes, 
\begin{equation} 
  {\rm det} < \delta U_j | L | \delta U_i > = 0. \label{det}  
\end{equation} 
The solution to the perturbation equations is found by successively  
solving the  
perturbed field equations for a trial function of the form given by 
equation (\ref{dU_exp})  
and then evaluating the 
determinant of equation (\ref{innerproduct}) 
for a sequence of stars with increasing 
rotation rate until the determinant's value passes through zero.  
The star for which the 
determinant is exactly zero, has a zero-frequency (neutral) $f$-mode, 
indicating the onset of the gravitational-radiation driven instability in this 
mode. 
  
Stergioulas \& Friedman (1998), also found that neutral $f$-modes can 
be determined with high accuracy (less than $1\%$ error) 
in an approximate gauge, in which only 
two perturbed field equations need to be solved, allowing a larger 
number of grid points to be used. The approximate gauge is defined by 
equations  (\ref{g1})-(\ref{g4}) supplemented by the 
approximations 
\begin{equation} 
\frac{h_{tt}}{g_{tt}+2\omega g_{t\phi}} = \frac{h_{rr}}{g_{rr}} =  
 \frac{h_{\theta \theta}}{g_{\theta \theta}}, \label{h1} 
\end{equation} 
and 
\begin{equation} 
h_{t \theta} = h_{r \phi}=0. \label{h2} 
\end{equation} 
Equation (\ref{h1}) enforces a similar relation between 
the diagonal components 
of $h_{ab}$, as in the Newtonian limit, while equation (\ref{h2}) 
essentially ignores 
the axial contribution to the metric perturbation (the axial contribution to 
the fluid velocity perturbation is retained).  
All results in the present paper will be obtained using the approximate 
gauge. Eight basis functions are used in equation (\ref{dU_exp}), corresponding 
to  $j=0..3$  and $k=0,1$. 

\section{Improved Numerical Implementation} 
\label{s:improved} 
 
In SF, a highly accurate finite-difference scheme was used for the 
angular variable, that allowed the use of only a few angular spokes. 
This finite-difference scheme requires the solution to be a very 
smooth function and it gave accurate results for relativistic 
polytropes of index $N \geq 1.0$. Realistic EOSs have, 
however, a stiff interior and a sudden drop in density near the 
surface. Hence, the finite difference scheme used in SF would suffer 
from the Gibbs phenomenon at the surface of the star, if it were  
applied to realistic EOSs. This would result in an error of several 
percent in the determination of neutral modes for these equations 
of state. 
 
Here, we use for the angular variable the same standard three-point 
finite difference scheme as used for the radial variable in SF.  With 
a fine enough grid, the density distribution near the surface is 
resolved accurately (see section \ref{s:eos}). 
  
Another improvement is the use of Ridder's method (see e.g. Press et. al. 
1992) for locating the 
exact point along a sequence of rotating stars, where the determinant 
goes through zero. In SF, linear interpolation between two nearby stars 
was used.

\section{Equations of State} 
\label{s:eos} 
 
The critical neutron star models for a set of EOSs were 
computed using the method described in the preceding section. Four 
realistic EOSs, A, C, L and WFF3, spanning a wide range 
of stiffness, were selected.  Equations of state A, C and L are 
labeled as in Arnett \& Bowers (1977).  Equation of state A is one of the softest 
EOSs allowing a nonrotating $1.4 M_\odot$ neutron star. 
Equation of state C has intermediate stiffness, while EOS L is one of the stiffest 
realistic EOSs.  Equation of state WFF3 (UV14+TNI in Wiringa, Fiks, 
\& Fabrocini 1988) is a 
modern EOS. At lower densities we match it to EOS FPS (Lorenz, Ravenhall, 
\& Pethick 1993), which accurately describes the crust of a neutron star (see 
Nozawa et al. 1998 for more details on the EOSs). 
 
\begin{figure}[t] 
\resizebox{\hsize}{10.5cm}{\includegraphics{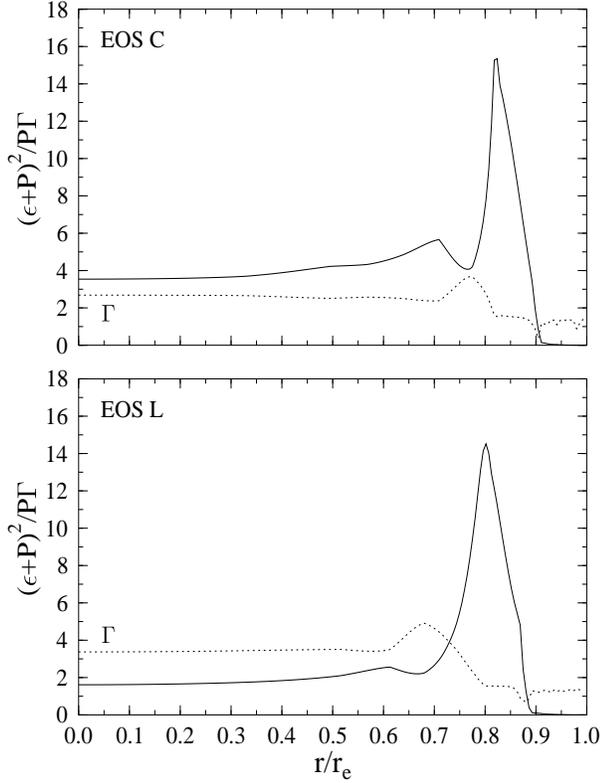}} 
\caption{Behaviour of $(\epsilon+P)^2/P\Gamma$ (solid line) and $\Gamma$  
  (dotted line) in the equatorial plane, as a function of coordinate 
  radius ($r_e$ is the coordinate radius at the equator). A $1.44 
  M_\odot$ EOS C model (upper panel) and a $1.38 M_\odot$ EOS L model 
  (lower panel), belonging to the $l=m=2$ neutral mode sequence are 
  shown. The units of the vertical axis are explained in the text. 
  Note that the divergence of $(\epsilon+P)^2/P\Gamma$ at the surface 
  is avoided by a softening of the EOS ($\Gamma$ becomes less than 
  2.0).}
\label{fig1} 
\end{figure}

All realistic EOSs examined have an adiabatic index 
$\Gamma$, which is larger than 2.0 at the center and for most of the 
interior of the star. Thus, the equilibrium models are similar to 
stiff polytropic models with index $N=1/(\Gamma-1)<1.0$.  For such 
polytropic models, the Eulerian perturbation in the energy density 
diverges at the surface. This poses a potential threat to our 
numerical scheme since the integrand of equation (\ref{innerproduct}) 
depends on $\delta \epsilon$ (cf. eq. (C24) of SF). Although the 
integral (\ref{innerproduct}) is formally finite, the divergence of 
$\delta \epsilon$ at the star's surface would make it difficult to 
accurately evaluate the integral. Skinner and Lindblom (1996) avoided 
this problem by using analytic expressions for the divergence of 
$\delta \epsilon$  in the case of Newtonian polytropes. In the 
realistic EOSs which we examined, this problem is not encountered 
because the EOSs soften near the star's surface.  In Figure \ref{fig1}, we 
plot the expression $(\epsilon+P)^2/P\Gamma$, which is proportional to 
$\delta \epsilon$, near the surface and the adiabatic index $\Gamma$, 
as a function of radial coordinate distance (in the equatorial plane) 
for representative equilibrium models constructed with EOSs C and L. 
As can be seen, the adiabatic index becomes less than $\Gamma=2.0$ 
near the surface, so that $\delta \epsilon$ goes to zero, as  occurs
 in soft 
polytropes of index $N>1.0$. With enough grid-points in the radial 
direction, this change in $\delta \epsilon$ can be resolved 
accurately. Note that $(\epsilon+P)^2/P\Gamma$ has a maximum at {\em 
  exactly} the points in the interior of the star, where the adiabatic 
index becomes $\Gamma=2.0$. The vertical axis in Figure \ref{fig1} is 
dimensionless (we set $c=G=1$ and the length scale equal to $c/\sqrt{G 
  \epsilon_0}$, where $\epsilon_0=10^{15} {\rm g/cm}^3$). 
 
The critical curves we obtain for the four realistic EOSs are also 
representative of the critical curves that one would obtain for stiff 
polytropes, if one  correctly handled the divergence of $\delta \epsilon$ at 
the surface or matched a soft polytropic surface to a stiff 
polytropic interior. For example, models constructed with EOS C are 
roughly similar to polytropic models of index $N\sim 0.7$, while 
models constructed with EOS L are roughly similar to polytropic models 
of index $N \sim 0.5$.

\begin{figure}[t] 
\resizebox{\hsize}{10.5cm}{\includegraphics{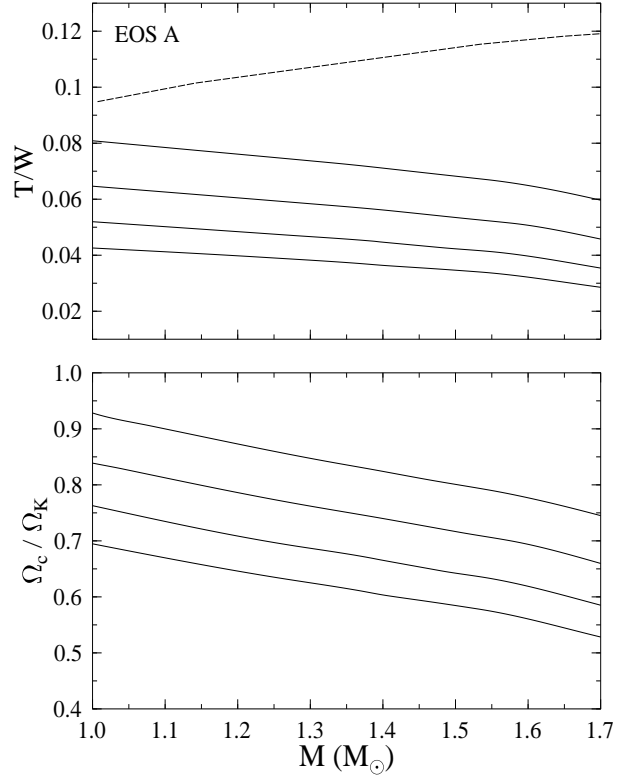}} 
\caption{Neutral mode sequences for EOS A. Shown are the ratio of rotational 
  to gravitational energy $T/W$ (upper panel) and the ratio of the 
  critical angular velocity $\Omega_c$ to the angular velocity at the 
  mass-shedding limit for uniform rotation (lower panel) as a function 
  of gravitational mass. The solid curves are the neutral mode 
  sequences for $l=m=2, 3, 4$ and 5 (from top to bottom), while the 
  dashed curve in the upper panel corresponds to the mass-shedding 
  limit for uniform rotation.}
\label{fig2} 
\end{figure} 
 
\begin{figure}[t] 
\resizebox{\hsize}{10.5cm}{\includegraphics{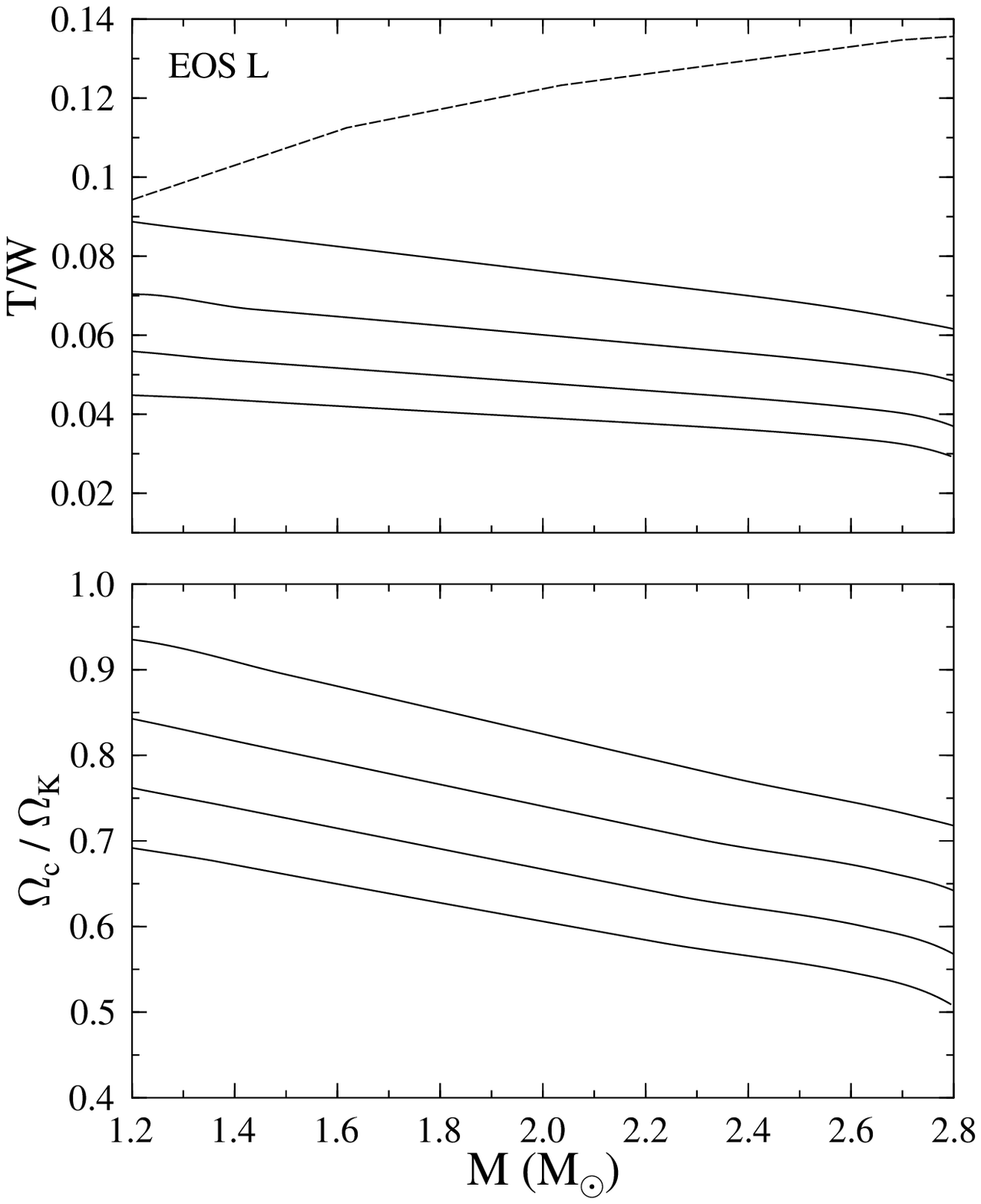}} 
\caption{Same as Figure 2 but for EOS L.}
\label{fig3} 
\end{figure} 
 
\begin{figure}[t] 
\resizebox{\hsize}{10.5cm}{\includegraphics{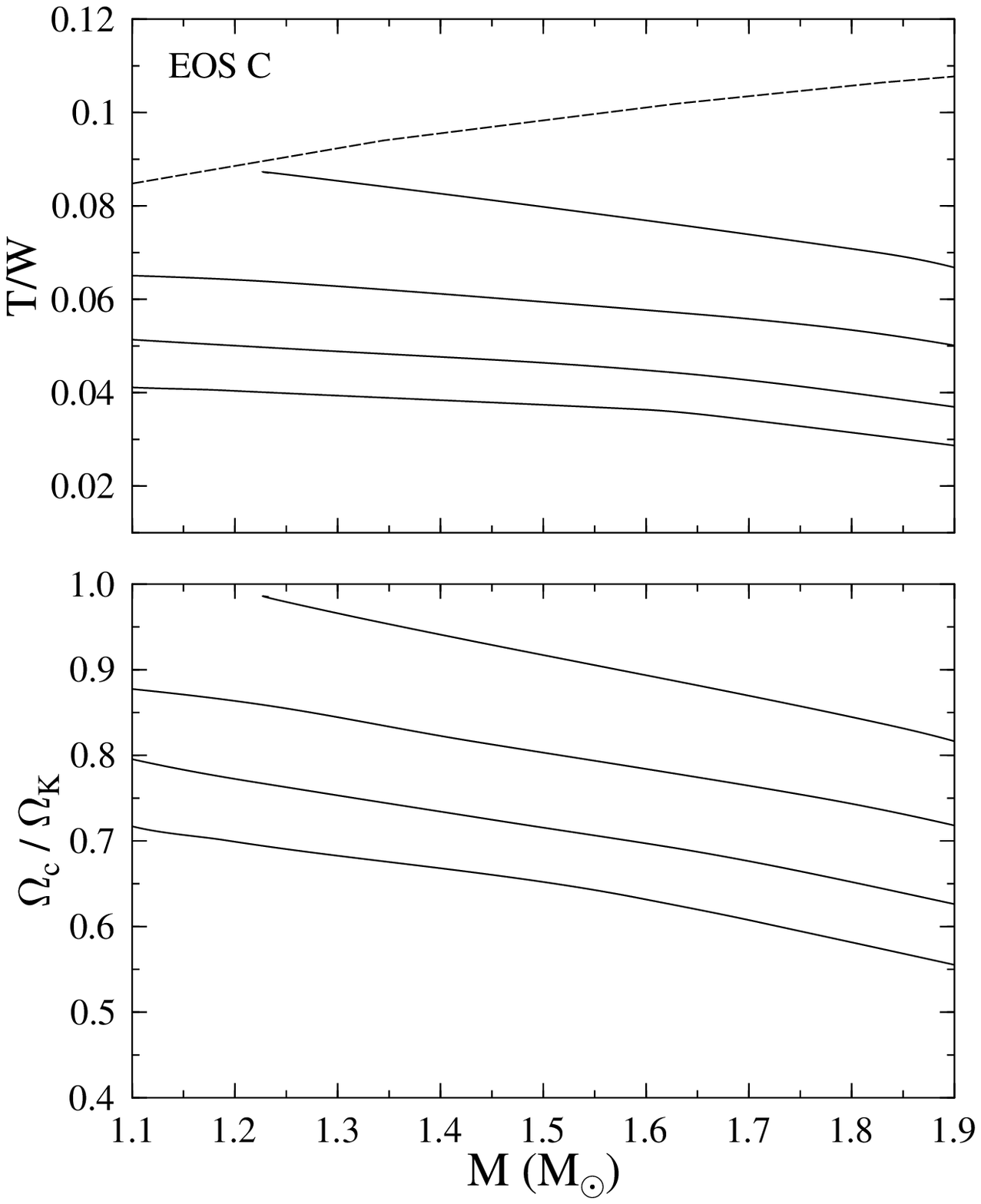}} 
\caption{Same as Figure 2 but for EOS C.} 
\label{fig4}
\end{figure} 
 
\begin{figure}[t] 
\resizebox{\hsize}{10.5cm}{\includegraphics{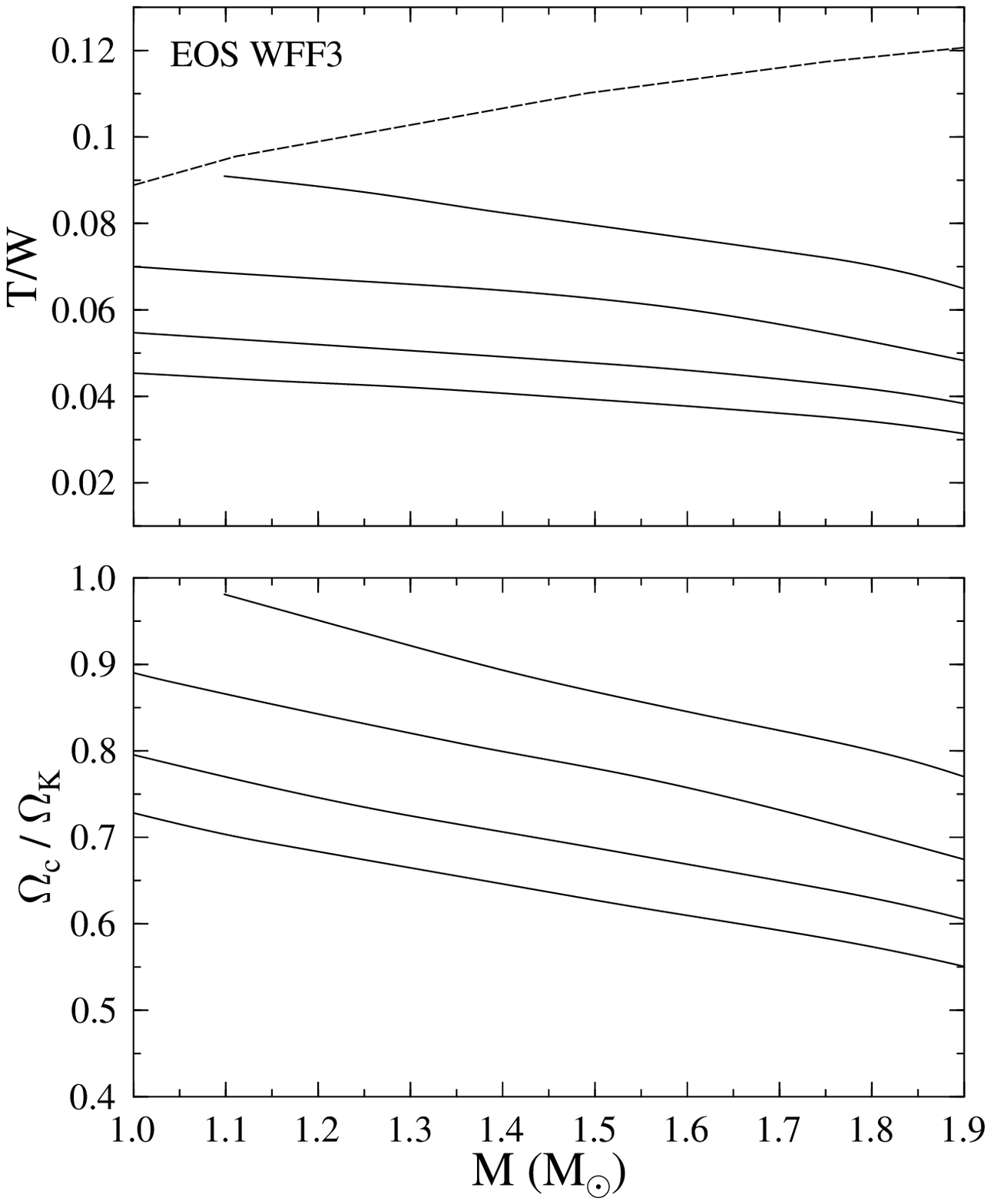}} 
\caption{Same as Figure 2 but for EOS WFF3.} 
\label{fig5}
\end{figure} 

\section{Critical curves for realistic equations  of state} 
\label{s:results} 
Results for each equation of state are summarized in Tables \ref{table_a} 
and \ref{table_c} and Figures \ref{fig2} to \ref{fig5}.  
The tables list equilibrium properties of the 
critical stars for a few selected examples. For each value of $m$, 
three stars were selected: a low mass star, a star with mass close to 
$1.4 M_\odot$ and a star close to the maximum mass stable star along 
each neutral mode sequence.  The following quantities are displayed in 
the tables: 
\begin{description} 
\item $\epsilon_c$ central energy density 
\item $T/W$ ratio of rotational energy to gravitational potential 
  energy 
\item $\Omega_c$ critical angular velocity 
\item $\Omega_c/\Omega_K$ ratio of critical angular velocity 
  $\Omega_c$ to the mass-shedding limit $\Omega_K$ at same central 
  energy density 
\item $M$ gravitational mass 
\item $M_0$ rest mass 
\item $R$ equatorial circumferential radius 
\end{description} 
The $l=m=2$ bar mode has the fastest growth time and will be the most 
efficient mode for the emission of gravitational radiation. For $1.4 
M_\odot$ stars, the bar mode is unstable for $\Omega/\Omega_K > 0.83$ 
for the softest EOS A and for $\Omega/\Omega_K > 0.93$ for EOS 
  C. This corresponds to critical spin periods of 0.8 ms and 1.1 ms 
for EOSs A and C respectively. For maximum mass stars, the bar mode is 
unstable for $\Omega/\Omega_K > 0.69$ at $M=2.9 M_\odot$ for the 
stiffest EOS L and $\Omega/\Omega_K > 0.77$ at $M=2.0 M_\odot$ for EOS 
C. 
 
In terms of $T/W$, the $l=m=2$ mode becomes unstable at $T/W \sim 
0.071 - 0.086$ for $1.4 M_\odot$ stars and at $T/W \sim 0.06$ for the 
maximum mass along the neutral mode sequence. The latter value is 
surprisingly insensitive to the equation of state. In fact, the 
$l=m=2$ neutral mode sequence can be approximated by the following 
linear empirical formula, 
\begin{equation} 
  \left( T/W \right)_2 = 0.115 -0.048 \frac{M}{M_{\rm max}^{\rm sph}},  
\label{emp} 
\end{equation} 
where $M_{\rm max}^{\rm sph}$ is the maximum mass for a spherical star 
allowed by a given equation of state. The empirical formula has an 
accuracy of roughly $4\%-6\%$ for all values of $M$, except for stars 
near the axisymmetric stability limit, that is, near the maximum mass 
along the neutral mode sequence, where it is somewhat larger, 
(the $T/W$ vs. $M$ curves deviate somewhat from linearity, as 
can be seen in figs. 2 to 5).  While in the Newtonian limit the 
  bar mode becomes unstable at $T/W \sim 0.14$, in realistic $1.4 
  M_\odot$ neutron stars the onset of the bar mode instability is at 
  roughly $1/2$ to $2/3$ the Newtonian estimate of $T/W$. 
 
The critical curves for the $l=m=3,4$ and 5 modes appear at 
successively lower rotation rates. All critical curves for all 
  EOSs are nearly linear in gravitational mass and  similar empirical 
formulae for these modes can also be written, as in equation (\ref{emp}). 
 
The obtained critical curves assume a perfect fluid. We expect that 
including the effects of viscosity will raise the critical angular 
velocities by a few percent in the $10^9$K to $10^{10}$ K temperature 
window. The strengthening of the instability by relativistic effects 
will also widen the temperature window in which the instability will 
be active, as was already shown by post-Newtonian computations (Cutler 
\& Lindblom 1992, Lindblom 1995). 
 
By doubling the number of grid-points in both directions and from 
comparisons with the Newtonian limit and with the results in SF, we 
estimate the accuracy of our present results to be at the $1\%$ 
level. Increasing the number of basis functions in equation (\ref{dU_exp})
 to more than  eight, did not affect the critical curves by more than $1\%$.  
 
\section{Discussion} 
 
We find that for a wide range of realistic EOSs the polar
$l=m=2$ bar mode is unstable to the emission of gravitational waves 
in newly-born $1.4 M_\odot$ neutron stars, rotating close to 
the Kepler limit, until their angular velocity falls below $83\% 
-93\%$ of the Keplerian value. The recent observation of the fastest 
rotating  young pulsar in the supernova remnant N157B (Marshall et 
al. 1998) suggests that a fraction of neutron  stars born in 
supernovae are born with very large initial rotational energy.  If 
  some neutron stars are born in accretion-induced collapse of white 
  dwarfs (Friedman 1983), they are also expected to have a large initial spin.  Since 
it is initially very hot and differentially rotating, a proto-neutron 
star can even be born with an angular velocity exceeding the 
mass-shedding limit for uniformly rotating stars.  As the star cools 
and passes through the temperature window of $10^9$K to $10^{10}$ K, 
the nonaxisymmetric bar mode, driven by gravitational radiation, will 
grow and the star will lose angular momentum by the emission of 
gravitational waves. Within a short time, the star will have slowed 
down enough that the bar mode will become stable again. During this 
first phase the $r$-mode instability will also be operating. The 
$r$-mode instability will then continue to slow down the star, until 
the star reaches a period of  roughly $6-9$ ms, when the 
instability will be damped by viscosity. 
 
The above picture assumes that the star cools through the standard 
modified URCA cooling scenario.  If instead neutron stars cool very 
rapidly through e.g. the direct URCA process, then the instability in 
$f$-modes may not have enough time to grow significantly and 
the rotational evolution of the star will only be affected by the 
$r$-mode instability. 
 
The CFS-instability in the bar $f$-mode appears to be a good source of 
detectable  continuous gravitational waves.  Lai \& Shapiro 
(1995) have studied the development of the $f$-mode instability using 
Newtonian ellipsoidal models of rotating stars (Lai, Rasio, \& Shapiro 1993, 
1994). They consider the case where a neutron star is created in a 
core collapse with large initial angular momentum. After a brief 
dynamical phase, the proto-neutron star becomes axisymmetric but 
secularly unstable. The instability deforms the star into a 
nonaxisymmetric configuration via the $l=m=2$ bar mode.  As the star 
slows down, the frequency of the gravitational waves sweeps downward 
from a few hundred Hz to zero, passing through LIGO's ideal 
sensitivity band. A rough estimate of the wave amplitude shows that, 
at $\sim 100$Hz, the gravitational waves from the CFS-instability 
could be detected out to the distance of 30 Mpc by LIGO or VIRGO and 
to 140Mpc by the advanced LIGO detector. This result is very 
promising, especially since for relativistic stars the instability 
will be stronger than in the Newtonian computations. 
 
Another astrophysical situation in which the instability may have the 
opportunity to grow is after the merger of two neutron stars in a 
binary coalescence.  Recently, Baumgarte \& Shapiro (1998) studied the 
case in which the merged neutron star is unstable to collapse, but has 
more angular momentum than required to collapse to a Kerr black hole. 
They find that neutrino emission is inefficient for shedding the 
excess angular momentum of the neutron star and suggest that this can 
happen through the growth of the gravitational radiation driven bar 
$f$-mode. We expect the gravitational waves from the instability in 
these high mass ($M>2.8 M_\odot$) merged neutron stars to be 
especially strong and a detailed, fully relativistic study is needed. 
 
The computation of quasi-normal finite-frequency modes of rapidly 
rotating relativistic stars is a more difficult problem than the 
neutral-mode calculation presented in this paper. The main difficulty 
is in applying the boundary conditions at infinity. Lindblom, Mendell 
and Ipser (1997) have recently proposed an approximate near-zone 
boundary condition which appears to be a promising approach for 
solving for the complex eigen-frequencies.  We plan to incorporate the 
near-zone boundary conditions into the SF method to allow the 
approximate computation of frequencies and growth times of the 
quasi-normal modes with reasonable accuracy.

\acknowledgments  
 
It is a pleasure to thank John Friedman for helpful discussions and 
Nils Andersson for a critical reading of the manuscript.  
We also thank S. Yoshida and Y. Eriguchi for providing us with graphs of 
critical curves for two realistic equations of state, obtained in the 
Cowling approximation, before their publication.  N.S. and S.M.M. 
acknowledge the generous hospitality of the Max-Planck-Institute for 
Gravitational Physics (Albert-Einstein-Institute) in Potsdam, 
Germany. This work was supported in part by NSF grant PHY 95-07740 and 
by NSERC of Canada.

%
%
%
%
%
%

\begin{deluxetable}{lccccccc} 
\tablecaption{Neutral mode sequences for EOS A and EOS L \label{table_a}} 
\tablewidth{0pt} 
\tablehead{ 
\colhead{} & 
\colhead{$\epsilon_c$}           & \colhead{$T/W$}      & 
\colhead{$\Omega_c$}          & \colhead{$\Omega_c/\Omega_K$}  & 
\colhead{$M$}         & \colhead{$M_0$}    & 
\colhead{$R$} \nl 
 \colhead{} & 
\colhead{$(\times 10^{15}{\rm g/cm}^3)$}           & \colhead{}      & 
\colhead{$(\times 10^{3} {\rm  s}^{-1})$}          & \colhead{}  & 
\colhead{($M_\odot$)}         & \colhead{($M_\odot$)}    & 
\colhead{(km)} } 
\tablecolumns{8} 
\startdata 
\cutinhead{EOS A}  
$m=2$ & 1.00 & 0.082 & 6.58 & 0.94 & 0.97 & 1.04 & 12.4   \nl 
 & 1.50 & 0.071 & 7.46 & 0.83 & 1.40 & 1.55 & 11.2   \nl 
 & 3.20 & 0.056 & 9.25 & 0.72 & 1.76 & 2.06 &  9.4   \nl 
\tablevspace{0.3cm} 
$m=3$ &1.00 & 0.066 &  6.02 & 0.86 & 0.94 & 1.00 & 11.7  \nl 
 & 1.60 & 0.056 &  6.90 & 0.74 & 1.41 & 1.58 & 10.7  \nl 
 & 3.20 & 0.044 &  8.28 & 0.65 & 1.73 & 2.03 & 9.3   \nl 
\tablevspace{0.3cm} 
$m=4$ &1.00  & 0.054 &  5.50 & 0.79 & 0.91 & 0.97 & 11.3   \nl 
 & 1.62 & 0.045 &  6.28 & 0.67 & 1.40 & 1.57 & 10.5   \nl 
 & 3.20 & 0.035 &  7.42 & 0.58 & 1.71 & 2.00 & 9.2   \nl 
\tablevspace{0.3cm} 
$m=5$ &1.00  & 0.044 &  5.04 & 0.72 & 0.89 & 0.95 & 11.0   \nl 
 & 1.68 & 0.036 &  5.80 & 0.60 & 1.41 & 1.58 & 10.3   \nl 
 & 3.20 & 0.029 &  6.76 & 0.53 & 1.70 & 1.99 & 9.1   \nl 
\cutinhead{EOS L}  
$m=2$ &0.35 & 0.089 & 4.08 & 0.96 & 1.20 & 1.27 & 18.9 \\ 
 & 0.38 & 0.086 & 4.20 & 0.91 & 1.38 & 1.48 & 18.3 \\ 
 & 1.20 & 0.057 & 5.67 & 0.69 & 2.90 & 3.43 & 15.1 \\ 
\tablevspace{0.3cm}                                                                          
$m=3$ &0.35 & 0.070 & 3.70 & 0.87 & 1.14 & 1.21 & 17.5 \\ 
 & 0.40 & 0.067 & 3.88 & 0.81 & 1.44 & 1.54 & 17.2 \\ 
 & 1.20 & 0.045 & 5.10 & 0.62 &  2.85 & 3.38 & 14.9 \\ 
\tablevspace{0.3cm}                                                                                
$m=4$ &0.35 & 0.056 & 3.38 & 0.79 & 1.10 & 1.16 & 16.8 \\ 
 & 0.40 & 0.054 & 3.53 & 0.74 & 1.39 & 1.50 & 16.6 \\ 
 & 1.20 & 0.036 & 4.58 & 0.56 & 2.82 & 3.34 & 14.7 \\ 
\tablevspace{0.3cm}                                                                                 
$m=5$ &0.35 & 0.046 & 3.09 & 0.73 & 1.08 & 1.13 & 16.3 \\ 
 & 0.40 & 0.044 & 3.23 & 0.68 & 1.36 & 1.46 & 16.3 \\ 
 & 1.20 & 0.029 & 4.16 & 0.51 & 2.79 & 3.31 & 14.6 \\ 
\enddata 
\end{deluxetable} 
 
\begin{deluxetable}{lccccccc} 
\tablecaption{Neutral mode sequences for EOS C and EOS WFF3 \label{table_c}} 
\tablewidth{0pt} 
\tablehead{ 
\colhead{} & 
\colhead{$\epsilon_c$}           & \colhead{$T/W$}      & 
\colhead{$\Omega_c$}          & \colhead{$\Omega_c/\Omega_K$}  & 
\colhead{$M$}         & \colhead{$M_0$}    & 
\colhead{$R$} \nl 
 \colhead{} & 
\colhead{$(\times 10^{15}{\rm g/cm}^3)$}           & \colhead{}      & 
\colhead{$(\times 10^{3} {\rm  s}^{-1})$}          & \colhead{}  & 
\colhead{($M_\odot$)}         & \colhead{($M_\odot$)}    & 
\colhead{(km)} } 
\tablecolumns{8} 
\startdata 
\cutinhead{EOS C}  
$m=2$ & 0.74 & 0.087 & 5.48 & 0.99 & 1.23 & 1.32 & 16.4 \\ 
 & 0.90 &  0.082 & 5.92 & 0.93 & 1.44& 1.56 & 14.9 \\ 
 & 2.50 &  0.059 & 8.13 & 0.77 & 2.00 & 2.30 & 11.2 \\ 
\tablevspace{0.3cm}                       
$m=3$ &0.70 &  0.066 & 4.75 & 0.89 & 1.11 & 1.18 & 14.9 \\ 
 & 0.95 &  0.061 & 5.38 & 0.82 &1.43& 1.56 & 13.8 \\ 
 & 2.50 &  0.046 & 7.30 & 0.69 & 1.96 & 2.26 & 11.0 \\  
\tablevspace{0.3cm}                                                        
$m=4$ &0.70 &  0.052 & 4.29 & 0.80 & 1.07 & 1.14 & 14.2 \\ 
 & 1.00 &  0.047 & 4.94 & 0.73 & 1.45 & 1.58 & 13.2 \\ 
 & 2.50 &  0.036 & 6.48 & 0.62 & 1.94 & 2.23 & 10.8 \\  
\tablevspace{0.3cm}                                                                 
$m=5$ &0.70 &  0.042 & 3.91 & 0.73 &  1.05 & 1.11 & 13.8 \\ 
 & 1.00 &  0.038 & 4.46 & 0.65 & 1.42 & 1.55 & 12.9 \\  
 & 2.50 &  0.028 & 5.79 & 0.55 & 1.92 & 2.20 & 10.7 \\ 
\cutinhead{EOS WFF3}  
$m=2$ &0.80 & 0.091 & 6.09 & 0.98 & 1.10 & 1.17 & 14.6 \\  
& 1.00 & 0.083 & 6.56 & 0.89 & 1.40 & 1.53 & 13.2 \\  
& 2.50 & 0.059 & 8.33 & 0.74& 1.98 & 2.31 & 10.7 \\   
\tablevspace{0.3cm}                                  
$m=3$ &  0.70 & 0.072 & 5.17 & 0.94 & 0.86 & 0.90 & 13.9 \\   
& 1.05 & 0.063 & 5.99 & 0.79 & 1.40 & 1.55 & 12.5 \\   
& 2.50 & 0.046 & 7.44 & 0.66 & 1.95 & 2.27 & 10.5 \\   
\tablevspace{0.3cm}                                         
$m=4$  &0.70 & 0.057 & 4.68 & 0.85 & 0.82 & 0.87 & 13.0 \\   
& 1.10 & 0.051 & 5.55 & 0.71 & 1.43 & 1.58 & 12.1 \\   
& 2.50 & 0.037 & 6.73 & 0.60 & 1.92 & 2.24 & 10.4 \\   
\tablevspace{0.3cm}                                
$m=5$  &0.70 & 0.047 & 4.31 & 0.79 & 0.80 & 0.85 & 12.6 \\   
& 1.10 & 0.042 & 5.09 & 0.65 & 1.40 & 1.55 & 11.8 \\   
& 2.50 & 0.031 & 6.18 & 0.55 & 1.91 & 2.23 & 10.3 \\   
\enddata 
\end{deluxetable} 
 

\begin{thebibliography}{DUM} 
\bibitem[]{} Andersson, N. 1998, ApJ, in press, preprint available  
             as gr-qc/9706075 
\bibitem[]{} Andersson, N., Kokkotas, K., \& Schutz, B. F.  
         1998, preprint, gr-qc/9805225 
\bibitem[]{} Arnett, W. D., \& Bowers, R. L. 1977, ApJ Sup., { 33}, 415 
\bibitem[]{} Baumgarte, T. W., \& Shapiro, S. L. 1998, ApJ, in press, 
 preprint available as astro-ph/9801294 
\bibitem[]{} Bonazzola, S., Frieben, J., \& Gourgoulhon, E. 1998, 
        A\&A 331, 280 
\bibitem[]{} Chandrasekhar, S. 1970, Phys.  
   Rev. Lett., { 24}, 611 
\bibitem[]{} Cutler, C. 1991, ApJ, 374, 248 
\bibitem[]{} Cutler, C. \& Lindblom, L. 1987, ApJ, 314, 234 
\bibitem[]{} Cutler, C. \& Lindblom, L. 1992, ApJ, 385, 630 
\bibitem[]{} Cutler, C., Lindblom, L., \& Splinter, R. J. 1990, ApJ, 363, 603 
\bibitem[]{} Friedman, J. L. 1978, Commun. Math. Phys., { 62}, 247 
\bibitem[]{} Friedman, J. L. 1983, Phys. Rev. Lett. 51, 11
\bibitem[]{} Friedman, J. L. \& Ipser, J. 1992, Phil. Trans. R. Soc. Lond. A, 
         340, 391 
\bibitem[]{} Friedman, J. L., \& Morsink S. M. 1998, ApJ, in press, 
preprint available as gr-qc/9706073 
\bibitem[]{} Friedman, J. L. \& Schutz, B. F.  
   1978, ApJ, { 222}, 281 
\bibitem[]{} Imamura, J. N., Friedman, J. L., \& Durisen, R. H. 1985, 
        ApJ, 294, 474 
\bibitem[]{} Ipser, J. R., \& Lindblom, L. 1990, ApJ, 355, 226 
\bibitem[]{} Ipser, J. R., \& Lindblom, L. 1991, ApJ, 373, 213 
\bibitem[]{} Ipser, J. R., \& Lindblom, L. 1992, ApJ, 389, 392 
\bibitem[]{} Lai, D., Rasio, F. A., \& Shapiro, S. L. 1993, ApJ Suppl., 88, 205  
\bibitem[]{} Lai, D., Rasio, F. A., \& Shapiro, S. L. 1994, ApJ, 373, 213 
\bibitem[]{} Lai, D., \& Shapiro, S. L. 1995, ApJ, 442, 259 
\bibitem[]{} Lindblom, L. 1995, ApJ, 438, 265 
\bibitem[]{} Lindblom, L., Owen, B. J., \& Morsink, S. M. 1998, 
        Phys. Rev. Lett., in press, preprint available as  
gr-qc/9803053 
\bibitem[]{} Lorenz, C. P., Ravenhall, D. G., \& Pethick, C. J. 1993, 
        Phys. Rev. Lett., 70, 379 
\bibitem[]{} Managan, R. A. 1985, ApJ, 294, 463 
\bibitem[]{} Marshall, F. E., Gotthelf, E. V., Zhang, W., 
         Middleditch, J., \& Wang, Q. D. 1998, ApJ, 499, L179 
\bibitem[]{} Nozawa, T., Stergioulas, N., Gourgoulhon, E., \& Eriguchi, Y.  
        1998, A\&A, in press,  
preprint available as gr-qc/9804048 
\bibitem[]{} Papaloizou, J. \& Pringle, J. E. 1978, 
        MNRAS, 182, 423 
\bibitem[]{} Press, W. H., Teukolsky, S., Vetterling, W. T., \& 
        Flannery, B. P. 1992, Numerical Recipes in C, Second Edition,  
        (Cambridge: Cambridge University Press)  
\bibitem[]{} Regge, T., \& Wheeler, J. A., 1957, Phys. Rev., 108, 1063 
\bibitem[]{} Shapiro, S. L., \& Zane, S. 1997, preprint, gr-qc/9711050 
\bibitem[]{} Skinner, D., \& Lindblom, L. 1996, ApJ, 461, 920 
\bibitem[]{} Stergioulas, N. 1996, The Structure and Stability of Rotating 
Relativistic Stars, PhD Thesis, University of Wisconsin-Milwaukee, Milwaukee, USA 
\bibitem[]{} Stergioulas, N., 1998, Rotating Stars in Relativity, to appear in 
Living Reviews in Relativity, http://www.livingreviews.org/, preprint available as 
gr-qc/9805012 
\bibitem[]{} Stergioulas, N.  \& Friedman, J. L. 1998, ApJ, 492, 301 
\bibitem[]{} Wiringa, R.B., Fiks, V. \& Fabrocini, A. 1988,  
        Phys. Rev. C 38, 1010 
\bibitem[]{} Yoshida, S., \& Eriguchi, Y. 1995, ApJ, 438, 830 
\bibitem[]{} Yoshida, S., \& Eriguchi, Y. 1997, ApJ, 490, 779 
\bibitem[]{} Yoshida, S., \& Eriguchi, Y. 1998, private communication
  
  
\end{thebibliography}
\end{document}